\documentclass[aps,pre,twocolumn,amsmath,amssymb,superscriptaddress]{revtex4}
\usepackage{bbm}
\usepackage{indentfirst}
\usepackage{graphicx}
\usepackage{amsmath,amssymb}
\usepackage{bpchem}

\usepackage{color}
\usepackage{tabularx}
\usepackage{multirow}
\usepackage{makecell}
\usepackage{array}
\usepackage{appendix}
\newcommand{\PreserveBackslash}[1]{\let\temp=\\#1\let\\=\temp}
\newcolumntype{C}[1]{>{\PreserveBackslash\centering}p{#1}}
\newcolumntype{R}[1]{>{\PreserveBackslash\raggedleft}p{#1}}
\newcolumntype{L}[1]{>{\PreserveBackslash\raggedright}p{#1}}
\begin{document}

\title{Generic criterion for explosive synchronization in heterogeneous phase oscillator populations}
\author{Can Xu}\email[]{xucan@hqu.edu.cn}
\affiliation{Institute of Systems Science and College of Information Science and Engineering, Huaqiao University, Xiamen 361021, China}
\author{Xuan Wang}
\affiliation{Institute of Systems Science and College of Information Science and Engineering, Huaqiao University, Xiamen 361021, China}
\author{Per Sebastian Skardal}\email[]{persebastian.skardal@trincoll.edu}
\affiliation{Department of Mathematics, Trinity College, Hartford, Connecticut 06106, USA}

\newcommand{\WARN}[1]{\textcolor{green}{#1}}
\newcommand{\NOTES}[1]{\textcolor{red}{#1}}
\begin{abstract}
Ordered and disordered behavior in large ensembles of coupled oscillators map to different functional states in a wide range of applications, e.g., active and resting states in the brain and stable and unstable power grid configurations. For this reason, explosive synchronization transitions, which facilitate fast, abrupt changes between these functional states, has recently seen significant interest from researchers. While previous work has identified properties of complex systems that support explosive synchronization, these investigations have been conducted largely on an ad-hoc basis. Here we provide an exact criterion for explosive synchronization in coupled phase oscillator ensembles by investigating the necessary relationship between dynamical and structural disorder.
This result provides a critical step towards untangling the intertwined properties of complex systems responsible for abrupt phase transitions and bistability in biological and engineered systems.
\end{abstract}

\maketitle
Synchronization is a universal phenomenon that emerges in both natural and man-made systems~\cite{Strogatz2003,Pikovsky2003}, and unraveling the underlying mechanism that give rise to emergent self-organizing behavior is of great importance in applications of physical, biological, technological, and social systems~\cite{Tom2017RMP}. The Kuramoto model of couple oscillators~\cite{Kuramoto1975}, originally proposed in 1975, and its many variants and generalizations, serve as particularly useful paradigms for studying collective behavior. In the classical Kuramoto model the interplay between coupling and heterogeneity of natural frequencies results in a continuous, non-equilibrium phase transition of the Kuramoto order parameter, delineating a partially ordered state from the completely disordered state~\cite{Strogatz2000PD,JA2005RMP}.

Beyond this continuous phase transition, discontinuous, i.e., explosive, phase transitions towards synchronization have also been observed and have since received ample attention from the nonlinear dynamics and complex networks communities~\cite{JGG2011prl,SB2016,RDS2019}. The most notable feature of such an abrupt phase transition is the presence of hysteresis that gives rise to a region of bistability between incoherent and coherent states. The implications of explosive synchronization transitions are significant as, most notably, they enable abrupt transitions between ordered and disordered states under relatively small perturbations or parameter changes. In neuroscience and power grid applications, where synchronized vs coherent states characterize active vs resting brain states or stable vs unstable energy configurations, characterizing an exact criterion for the emergence of explosive transitions could be critical for system function~\cite{EM2015prx,MA2018PRL,MR2012PRL}. Thus, there is a critical need to understand the impacts of network structures, coupling schemes, and adaptive patterns on the emergence of explosive synchronization in complex systems~\cite{DP2005PRE,EA2009PRE,JGG2011prl,ZY2014PRL,JP2013PRL,XZ2015PRL,SC2019PRX,AP2020PRL,XD2020PRL,SB2016PR,RD2019ADV,CK2021SA,ZW2019PRE,XC2019PREE,CW2021EPJB,WX2021PRE, Skardal2014PRE,Skardal2020CommPhys}. Despite substantial attention and observed behavior in many variations and extensions of the Kuramoto model, a mathematically rigorous condition for the origin of explosive synchronization in coupled phase oscillator populations remains lacking.

The aim of this Letter is to fill this critical gap by providing such an exact criterion for explosive synchronization transitions. We focus our attention on phase oscillators with quenched disorder in both the local dynamics and structure, characterized by heterogeneity in natural frequencies and individual coupling strengths, respectively. In particular, we explore the effects that correlations between dynamical and structural disorder have on the resulting collective dynamics. The emergence of an explosive synchronization transition, characterized by a hysteresis loop supporting bistability between synchronized and incoherent states, is established using a self-consistent approach.
We uncover the origin of such abrupt transitions by revealing the local bifurcation mechanism of the system near criticality. Specifically, explosive transitions emerge when the concavities of the natural frequency distribution the coupling strength function combine to satisfy a particular inequality. Our results establish a rigorous characterization of the interplay between structure and dynamics that leads to explosive synchronization dynamics.

We consider here a generalized Kuramoto model of the form
\begin{equation}\label{equ:01}
\dot{\theta_{i}}=\omega_{i}+\frac{K_{i}}{N}\sum_{j=1}^N  \sin(\theta_{j}-\theta_{i}),\quad i=1,...,N.
\end{equation}
Here, $\theta_{i}\in S^1$ denotes the phase of oscillator $i$, $\omega_{i}$ is its natural frequency chosen randomly from a prescribed distribution $g(\omega)$, N is the size of the system, and $K_{i}$ accounts for the heterogeneous interactions between phase oscillators~\cite{DI2013PRL,DI2014NC}. Thus, the state of the system is described by a flow on the N-torus $\mathbb{T}^N=(S^{1})^{N}$. To establish correlations between the oscillator's local dynamics and coupling, we set $K_{i}=Kf(\omega_{i})$ so that the constant $K\ge0$ controls the overall coupling strength and $f(\omega)\ge0$ characterizes correlations between dynamical and structural disorder~\cite{XU2016PRE}. In particular, $f(\omega)=1$ recovers the conventional Kuramoto model, while the choice of $f(\omega)$ as a step function with mixed signs represents the constrain-conformist model investigated in previous studies~\cite{HH2011PRL,HH2016PRE,HH2016CHAOS}. The model defined by Eq.~(\ref{equ:01}) has rich and diverse rhythmic dynamics depending upon $g(\omega)$ and $f(\omega)$~\cite{HB2016PRL,CX2018PRE,CX2019NJP,YJ2021CHAOS,CX2021PRR}.

The aim of this work is to uncover a generic condition for the occurrence of explosive synchronization induced by the correlated disorder described above. For this purpose we restrict our attention to the case where $g(\omega)$ and $f(\omega)$ are even functions, i.e., symmetric about their mean, which we may set to $\omega=0$ without loss of generality by entering an appropriate reference frame, and twice continuously differentiable at $\omega=0$, but make no further restrictions on $g(\omega)$ and $f(\omega)$. To quantify the overall degree of synchronization we use the complex order parameter $Z(t)=R(t)e^{i\Theta(t)}=N^{-1}\sum_{j=1}^N e^{i\theta_j(t)}$, for which the amplitude $R(t)\in[0,1]$ measures the coherence of the system, and the argument $\Theta(t)\in S^1$ locates the average phase of the population.

 We first illustrate an exact result giving a criterion for the emergence of explosive synchronization. To this aim, we focus on finding the equilibrium states assuming that the system Eq.~(\ref{equ:01}) evolves to a steady state in $t\rightarrow\infty$. More precisely, we assume that $R(t)$ approaches a constant value and $\Theta(t)$ processes at a constant angular velocity. In fact, by setting the mean frequency to $\omega=0$, we may assume that this angular velocity is zero and then by shifting initial conditions we may set $\Theta(t)=0$. Therefore, the governing equation Eq.~(\ref{equ:01}) can be rewritten using the mean-field,
\begin{equation}\label{equ:03}
\dot{\theta}=\omega-qf(\omega) \sin\theta,
\end{equation}
where $q=KR$ and the index $i$ is dropped in the continuum limit, $N\rightarrow\infty$.

Next, at steady-state the phase oscillators split into two distinct groups depending on their frequencies. Specifically, those satisfying $|\omega|<qf(\omega)$ become phase-locked, i.e., entrained by the mean-field, and ultimately converge to a state satisfying $\sin\theta_{\omega}=\omega /[{qf(\omega)}]$ and $\cos\theta_{\omega}=\sqrt{1-\sin^{2}\theta_{\omega}}$. On the other hand, those satisftying $|\omega|>qf(\omega)$ never come to rest and remain drifting on $S^1$ for all time. The order parameter is expressed as $Z=\langle e^{i\theta}\rangle_{l}+\langle e^{i\theta}\rangle_d$, where $\langle {\cdot}\rangle_l$ and $\langle {\cdot}\rangle_d$ represent the ensemble average over the locked and drifting populations, respectively. It can be shown that, due to the symmetry of $g$ and $f$, $\langle e^{i\theta}\rangle_{d}=\langle=0$ and $\sin{\theta}\rangle_{l}=0$, which implies that the drifting populations have no contribution to the order parameter, and the locked ones contribute to its real part only~\cite{RM2007NS}. Thus, the self-consistent equation describing the steady state of the system is $Z=R=\langle \cos{\theta}\rangle_{l}$, which can be reformulated in a simple form given by
\begin{equation}\label{equ:08}
\frac{1}{K}=H(q)=\frac{1}{q}\int_{|\omega|<qf(\omega)}\sqrt{1-\frac{\omega^2}{q^{2}f^{2}(\omega)}}g(\omega)d\omega.
\end{equation}

Setting $F(\omega)=\omega /{f(\omega)}$, and defining $x={F(\omega)}/{q}$, the function $H(q)$ may be simplified to
\begin{equation}\label{equ:10}
H(q)=\int_{-1}^{1}\sqrt{1-x^2}G(qx)dx,
\end{equation}
where $G(qx)$ is denoted the generalized distribution (virtual distribution)~\cite{GJ2020PRE} given by
\begin{equation}\label{equ:11}
G(qx)=\frac{g[F^{-1}(qx)]f[F^{-1}(qx)]}{1-qxf^{'}[F^{-1}(qx)]},
\end{equation}
$F^{-1}(qx)$ is the inverse function determined by $qx=F(\omega)$, and the apostrophe represents the derivative. The self-consistent equations Eq.~(\ref{equ:08})--(\ref{equ:11}) characterize the degree of synchronization at steady-state. Since the incoherent state corresponds to $q=0$, $G(0)=g(0)f(0)$, the critical coupling strength $K_c$ characterizing the onset of synchronization via the instability of the incoherent state may be identified by taking the limit $q\rightarrow0^{+}$, giving
\begin{equation}\label{equ:12}
K_c=\frac{2}{\pi g(0)f(0)}.
\end{equation}
Remarkably, Eq.~(\ref{equ:12}) is a straightforward generalization of the classical result obtained by Kuramoto~\cite{Kuramoto1975}, now with $f(0)$ modifying the onset of synchronization.

Next, to identify an explosive synchronization transition, which is marked by a subcritical bifurcation at $K=K_c$ (contrasting a continuous transition, which is marked by a supercritical one), we let $K=K_c+\delta K$ and $q=0+\delta q$, with $\delta K,\delta q\ll1$. Expanding both sides of Eq.~(\ref{equ:08}) yields
\begin{equation}\label{equ:13}
\frac{1}{K_c}-\frac{\delta K}{K_{c}^2}=H(0)+H'(0)\delta q+\frac{1}{2}H''(0)\delta q^2.
\end{equation}
The first terms on either side of Eq.~(\ref{equ:13}) balance due to Eq.~(\ref{equ:12}), and the second term on the right hand side of Eq.~(\ref{equ:13}) vanishes due to the symmetry of $G(qx)$. Using the relation $\delta q^2\approx K^2_{c}\delta R^2$, we obtain that at criticality the order parameter is given by~\cite{XC2020PRE},
\begin{equation}\label{equ:15}
\delta R=\sqrt{\frac{-2\delta K}{H''(0)K^{4}_c}}.
\end{equation}
We conclude from Eq.~(\ref{equ:15}) that $H''(0)>0$ is the desired condition for subcriticality, i.e., explosive synchronization, where the phase transition from incoherence to synchronization is abrupt. On the other hand, when $H''(0)<0$ the bifurcation supercritical, yielding a continuous phase transition. In fact, we have $
H''(0)=\int_{-1}^{1}x^{2}\sqrt{1-x^2}G''(0)dx=\frac{\pi}{8}G''(0)$, so the value $G''(0)$ suffices to dictate the nature of the transition at $K=K_c$.

To obtain a more explicit condition for explosive synchronization, we expand $g(\omega)$ and $f(\omega)$ at the mean$\omega=0$, i.e., $g(\omega)=g(0)+\frac{g''(0)}{2}\omega^2+\mathcal{O}(\omega^4)$ and $f(\omega)=f(0)+\frac{f''(0)}{2}\omega^2+\mathcal{O}(\omega^4)$. Without loss of generality, $f(0)$ can be set to one by rescaling the time $t$. To leading order we then have $F^{-1}(qx)=({1-\sqrt{1-2f''(0)q^{2}x^2}})/[{f''(0)qx}]$, with $0<qx\ll1$. The generalized distribution $G(qx)$ near $qx=0$ is then given by, to leading order,
\begin{equation}\label{equ:18}
G(qx)=\frac{\omega[g(0)+2^{-1}g''(0)]}{qx[1-qxf''(0)\omega]}\Big|_{\omega=F^{-1}(qx)}.
\end{equation}
Using $F^{-1}(qx)$, we eventually obtain $G''(0)=g''(0)+3g(0)f''(0)$, yielding a universal condition for the onset of explosive synchronization given by
\begin{equation}\label{equ:19}
g''(0)+3g(0)f''(0)>0.
\end{equation}

\begin{figure}[t]
\centering
\includegraphics[width=1.0\columnwidth]{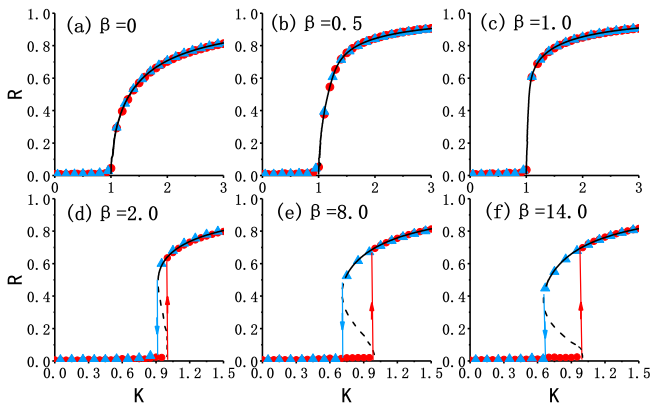}
\caption{ {\it Continuous vs explosive synchronization transitions.} The order parameter $R$ vs. coupling strength $K$ for $g(\omega)=\gamma/[\pi (\omega^2 + \gamma^2)]$ with $\gamma=0.5$ and $f(\omega)=2-e^{-\beta \omega^2}$, giving $\beta_c=4/3$. Panels (a)--(c) correspond to $\beta<\beta_c$, yielding continuous transitions, while panels (d)--(f) corresponds to $\beta>\beta_c$ exhibiting explosive transitions. Red and blue circles indicate results from numerical simulations with $N=10^5$ oscillators, adiabatically increasing and decreasing $K$, respectively. The solid (stable) and dashed (unstable) lines represent theoretical predictions, respectively.}%
 \label{fig:01}
 \end{figure}

Once the inequality in Eq.~(\ref{equ:19}) is satisfied, the phase transition towards synchronization is explosive. For example, if $g(\omega)={\gamma}/[{\pi}({\omega^2 +\gamma^2})]$ and $f(\omega)=2- e^{-\beta \omega^2}$ the critical parameter for the onset of explosive synchronization is $\beta_c =1/({3\gamma^2})$, above which the transition is abrupt. Using $\gamma=1/2$, yielding a critical value of $\beta_c=4/3$ we illustrate our main result in Fig.~\ref{fig:01}(a)--(f), plotting for $\beta=0$, $0.5$, $1$, $2$, $8$, and $14$ the synchronization transitions $R$ vs $K$. Results from direct simulations with $N=10^5$ oscillators are plotted in red and blue circles, indicating simulations where $K$ are adiabatically increased and decreased, respectively, and theoretical curves are plotted in black. Note that all simulations with $\beta<\beta_c$ (top row) yields continuous transitions, and those with $\beta>\beta_c$ (bottom row) yield explosive transitions. More generally, for the typical case of a unimodal frequency distribution with $g''(0)<0$, it follows that the transition to synchronization is explosive only for sufficiently positive $f''(0)$, specifically, $f''(0)>-g''(0)/[3g(0)]$. Note that, given the conditions on $f(\omega)$, this requires that there exists some open interval containing $\omega=0$ where $f(\omega)$ is both concave up as well as monotonically increasing as $|\omega|$ increases. Thus, in some interval near the mean frequency, explosive synchronization requires a sufficiently positive correlation between the natural frequencies and local coupling strengths.

For the remainder of this letter we present a rigorous bifurcation analysis of the onset of synchronization that occurs at $K=K_c$, putting our main result on a rigorous mathematical foundation. We begin by noting that in the continuum limit the governing equation Eq.~(\ref{equ:01}) is equivalent to the following continuity equation
\begin{equation}\label{equ:20}
\frac{\partial\rho}{\partial t}+\frac{\partial(\rho v)}{\partial \theta}=0
\end{equation}
with velocity field $v(\theta,\omega,t)=\omega+{Kf(\omega)}\mathrm{Im}(Ze^{-i\theta})$ and $\rho (\theta,\omega,t)$ representing the real, $2\pi$-periodic distribution describing the density of oscillators with phase $\theta$ and natural frequency $\omega$ at time $t$ whose $n^{\text{th}}$ Fourier coefficient of $\rho$ is defined by $\alpha_n=\int_0^{2\pi}e^{in\theta}\rho(\theta,\omega,t)d\theta$ for with $\alpha_0 =1$ and ${\alpha}_{-n}(\omega,t)={\bar{\alpha}}_{n}(\omega,t)$, where the over-bar denotes complex conjugate.

The Ott-Antonsen ansatz restricts the Fourier coefficients to a special form~\cite{OTT2008CHAOS,OTT2009CHAOS} defined by ${\alpha}_{n}(\omega,t)={\alpha}^{n}(\omega,t)$ for $n\ge1$, which is an attracting, invariant manifold of Eq.~(\ref{equ:20}). On this manifold, the low dimensional dynamics of the system evolve according to
\begin{equation}\label{equ:23}
\frac{d\alpha}{dt}=i\omega \alpha+\frac{Kf(\omega)}{2}(Z-\bar{Z}\alpha^2),
\end{equation}
which is closed with the order parameter $Z(t)$ defined by the integral
\begin{equation}\label{equ:24}
Z(t)=\int_{-\infty}^{+\infty}\alpha(\omega,t)g(\omega)d\omega=\mathcal{G}\alpha(\omega,t).
\end{equation}
(The integral operate $\mathcal{G}$ is introduced to ease notation.) The steady state solutions of the low dimensional dynamics in Eq.~(\ref{equ:23}) are expressed as a simple profile~\cite{OE2013PD},
\begin{equation}\label{equ:25}
 \alpha_s(\omega) =
\begin{cases}
\frac{i\omega+\sqrt{q^2 f^2 (\omega)-\omega^2}}{qf(\omega)}, &  |\omega|\le qf(\omega),\\
i \frac{\omega-\mathrm{sgn}(\omega)\sqrt{\omega^2-q^2 f^2 (\omega)}}{qf (\omega)},&  |\omega|>qf(\omega),
\end{cases}
\end{equation}
where $\mathrm{sgn}(\omega)$ is the sign function with respect to $\omega$, and the first and second terms corresponds to the phase-locked and drifting populations, respectively.

To analyze the stability of the steady solutions $\alpha(\omega)$ we consider small perturbations away from $\alpha_s(\omega)$, namely, $\alpha(\omega,t)=\alpha_s(\omega)+\varepsilon v(\omega,t)$, where $0<\varepsilon\ll1$ and $v(\omega,t)$ is the perturbation vector. Then, the order parameter may be written $Z(t)=R+\varepsilon \mathcal{G}v(\omega,t)$ and the linearized dynamics around $\alpha_s(\omega)$ is
\begin{equation}\label{equ:29}
\frac{dv}{dt}=\eta_{\omega}v+\frac{K}{2}(a_{\omega}\mathcal{G}v+b_{\omega}\mathcal{G}\bar{v}),
\end{equation}
where 
$\eta_{\omega}=i\omega -qf(\omega)\alpha_{s}(\omega)$, and the coefficients $a_{\omega}=f(\omega)$ , $b_{\omega}=-f(\omega)\alpha_s ^2 (\omega)$. By introducing $\mathbf{V}=(v,\bar{v})^T$ we transform Eq.~(\ref{equ:29}) to
\begin{equation}\label{equ:30}
 \frac{d\mathbf{V}}{dt}=\mathbf{M}\mathbf{V}+\frac{K}{2}Q\hat{{g}}\mathbf{V},
 \end{equation}
where
$\mathbf{M} = \left(
\begin{array}{cccc}
\eta_{\omega} & 0 \\
0 & \bar{\eta}_{\omega}\\
\end{array} \right)$, and
$\mathbf{Q}= \left(
\begin{array}{cccc}
a_{\omega} & b_{\omega} \\
\bar{b}_{\omega}  & \bar{a}_{\omega}\\
\end{array} \right)$. Using $d\mathbf{V}/dt=\lambda\mathbf{V}$ we obtain
\begin{equation}\label{equ:31}
 \mathbf{V}=\frac{K}{2}(\lambda\mathbf{I}-\mathbf{M})^{-1}\mathbf{Q}\mathcal{G}\mathbf{V},
 \end{equation}
where $\mathbf{I}$ is the unit matrix. Applying the integral operate $\mathcal{G}$ to both sides of Eq.~(\ref{equ:31}) leads to the following two dimensional linear equations
\begin{equation}\label{equ:32}
(\mathbf{I}-K\mathbf{J})\mathcal{G}\mathbf{V}=0,
 \end{equation}
with
 \begin{equation}\label{equ:33}
\mathbf{J}=\frac{1}{2}\mathcal{G}[(\lambda\mathbf{I}-\mathbf{M})^{-1}\mathbf{Q}].
 \end{equation}
A nontrivial solution of Eq.~(\ref{equ:32}) requires that $\det(\mathbf{I}-K\mathbf{J})=0$, from which we arrive at the pair of eigenvalue equations
 \begin{align}
\frac{1}{K}=h_c (\lambda)&=\frac{1}{2}\mathcal{G}\frac{(1-\alpha_{s}^2)f(\omega)}{\lambda-\eta_\omega},\label{equ:34}\\
\frac{1}{K}=h_s (\lambda)&=\frac{1}{2}\mathcal{G}\frac{(1+\alpha_{s}^2)f(\omega)}{\lambda-\eta_\omega},\label{equ:35}
 \end{align}
where $\lambda\in\mathbb{R}$ and $\lambda\neq \eta_\omega$.

Recall that, for the incoherent state $\alpha_s (\omega)=0$, $\eta_{\omega}=i\omega$, we have
\begin{equation}\label{equ:35+}
h_c (\lambda)=h_s (\lambda)=\frac{1}{2}\int_{-\infty}^{+\infty}\frac{\lambda f(\omega)g(\omega)}{\lambda^2 +\omega^2}d\omega.
 \end{equation}
In the limit $\lambda\rightarrow0^+$, the function $\lambda/(\lambda^2 +\omega^2)\rightarrow \pi \delta(0)$, yielding, $K\rightarrow K_c =2/[\pi g(0)f(0)]$, which coincides with onset of synchronization obtained previously. Thus, the phase transition towards synchronization takes place at the critical coupling strength $K_c$ at which point the incoherent state loses stability.

As for the partially synchronized state with $\alpha_s (\omega)\neq 0$, the sign of $\lambda$ controls their stability properties. It can be shown that $h_s(0)=q^{-1}\mathcal{G}\alpha_0 (\omega)$, which is exactly the self-consistent equation Eq.~(\ref{equ:08}). This implies that $\lambda =0$ is always a trivial root of Eq.~(\ref{equ:35}) originating form the rotational symmetry of Eq.~(\ref{equ:01}). However, for $\lambda\neq0$, we define $\Delta (\lambda)=h_c (\lambda)-K^{-1}$, which satisfies
 \begin{equation}\label{equ:36}
\Delta(0)=qH'(q).
 \end{equation}
To prove Eq.~(\ref{equ:36}) we note that
 \begin{equation}\label{equ:37}
qH'(q)=q[\frac{1}{q}\mathcal{G}\alpha_s (\omega)]'=\mathcal{G}[\frac{d\alpha_s (\omega)}{dq}]-\frac{1}{K}.
 \end{equation}
Following Eq.~(\ref{equ:25}), we have that, for $|\omega|<qf(\omega)$,
 \begin{equation}\label{equ:38}
\frac{d\alpha_s }{dq}=i\frac{\sin^2 {\theta}_\omega}{q\cos{\theta}_\omega},
 \end{equation}
and for $|\omega|>qf(\omega)$,
\begin{equation}\label{equ:39}
\frac{d\alpha_s }{dq}=i[\frac{f(\omega)\text{sgn}(\omega)}{\sqrt{\omega^2 -q^2 f^2 (\omega)}}-\frac{\omega-\text{sgn}(\omega)\sqrt{\omega^2 -q^2 f^2 (\omega)}}{q^2 f (\omega)}].
 \end{equation}
Putting all this together, we obtain
\begin{equation}\label{equ:40}
h_c (0)=\mathcal{G}\left(\frac{d\alpha_s }{dq}\right),
 \end{equation}
which completes the proof of Eq.~(\ref{equ:36}). In Ref.~\cite{CX2021PREE} it was shown that $\Delta (0)>0$ implies instability. By contrast, $\Delta(0)<0$ indicates that the corresponding stationary solution described by $H(q)$ remains asymptotically stable. According to Eq.~(\ref{equ:13}), $H'(q)\sim H''(0)q$ for $0<q\ll1$. Hence, for $H''(0)>0$ the associated branch of the bifurcating solutions is unstable, and in the limit $q\rightarrow0^+$ this unstable branch collides with the stable incoherent state, thereby creating a subcritical bifurcation at $K_c$, i.e., an explosive transition to synchronization.

Take together, we have fully characterized the dynamics of a generalized Kuramoto model, in which the interactions are deterministically correlated with the given oscillator's intrinsic frequency. We have revealed a universal condition that leads to the explosive synchronization via hysteresis and bistability. Furthermore, we have demonstrated the mathematical basis for such an abrupt transition by developing the local bifurcation theory of the incoherent state at onset. Our study could find applicability in better understanding the nonequilibrium phase transitions and related critical phenomenon in complex systems.

\section*{ACKNOWLEDGEMENTS}
CX and XW acknowledge support from the National Natural Science Foundation of China (Grants No. 11905068) and the Scientific Research Funds of Huaqiao University (Grant No. ZQN-810). PSS acknowledges support from the National Science Foundation (Grant No. MCB-2126177). We thank the support from Higher-Order Network Reading Group supported by the Save 2050 Programme jointly sponsored by Swarma Club and X-Order.

\end{document}